\documentclass[prl,aps,floats,float,twocolumn]{revtex4}
\usepackage{epsfig}
\begin{document}

\title{Absolute Frequency Measurements of the ${\rm Hg^+}$ and Ca Optical
Clock Transitions with a Femtosecond Laser}

\author{Th.~Udem$^*$, S.~A.~Diddams, K.~R.~Vogel, C.~W.~Oates,
        E.~A.~Curtis, W.~D.~Lee, W.~M.~Itano, R.~E.~Drullinger,
        J.~C.~Bergquist, and L.~Hollberg}
\affiliation{Time and Frequency Division, National Institute of
         Standards and Technology, \\325 Broadway, Boulder,
         Colorado 80305}



\begin{abstract}
The frequency comb created by a femtosecond mode-locked laser and
a microstructured fiber is used to phase coherently measure the
frequencies of both the ${\rm Hg^+}$ and Ca optical standards with
respect to the SI second as realized at NIST. We find the
transition frequencies to be
$f_{Hg}=1\;064\;721\;609\;899\;143(10)\;\mbox{Hz}$ and
$f_{Ca}=455\;986\;240\;494\;158(26)\;\mbox{Hz}$, respectively. In
addition to the unprecedented precision demonstrated here, this
work is the precursor to all-optical atomic clocks based on the
${\rm Hg^+}$ and Ca standards. Furthermore, when combined with
previous measurements, we find no time variations of these atomic
frequencies within the uncertainties of $\vert (\partial
f_{Ca}/\partial t)/f_{Ca} \vert \leq 8\times 10^{-14}\;{\rm
yr^{-1}}$, and $\vert (\partial f_{Hg}/\partial t)/f_{Hg} \vert
\leq 30\times 10^{-14}\;{\rm yr^{-1}}$.
\end{abstract}
\maketitle

\vspace{1cm}

Optical standards based on a single ion or a collection of
laser-cooled atoms are emerging as the most stable and accurate
frequency sources of any
sort~\cite{Rafac,Bernard99,Schnatz,Oates00,Indium}. However, because
of their high frequencies ($\sim500\;\mbox{THz}$), it has proven
difficult to count cycles as required for building an optical clock
and comparing to the cesium microwave standard. Only recently, a
reliable and convenient optical clockwork fast enough to count
optical oscillations has been
realized~\cite{Diddams,Jones,Holzwarth}. Here, we report an optical
clockwork based on a single femtosecond laser that phase coherently
divides down the visible radiation of the ${\rm Hg^+}$ and Ca
optical frequency standards to a countable radio frequency. By this
means we determine the absolute frequencies of these optical
transitions with unparalleled precision in terms of the SI second as
realized at NIST~\cite{Jefferts}. Indeed, for the ${\rm Hg^+}$
standard, the statistical uncertainty in the measurement is
essentially limited by our knowledge of the SI second at $\sim
2\times 10^{-15}$. The high precision and high demonstrated
stability of the standards\cite{Rafac,Oates00} combined with the
straightforward femtosecond-laser-based clockwork suggest ${\rm
Hg^+}$ and Ca as excellent references for future all-optical clocks.
Additionally, the comparison of atomic frequencies over time
provides constraints on the possible time variation of fundamental
constants. When combined with previous measurements, the current
level of precision allows us to place the tightest constraint yet on
the possible variation of optical frequencies with respect to the
cesium standard.

\begin{figure}[bthp]
\begin{center}
\includegraphics[width=8cm]{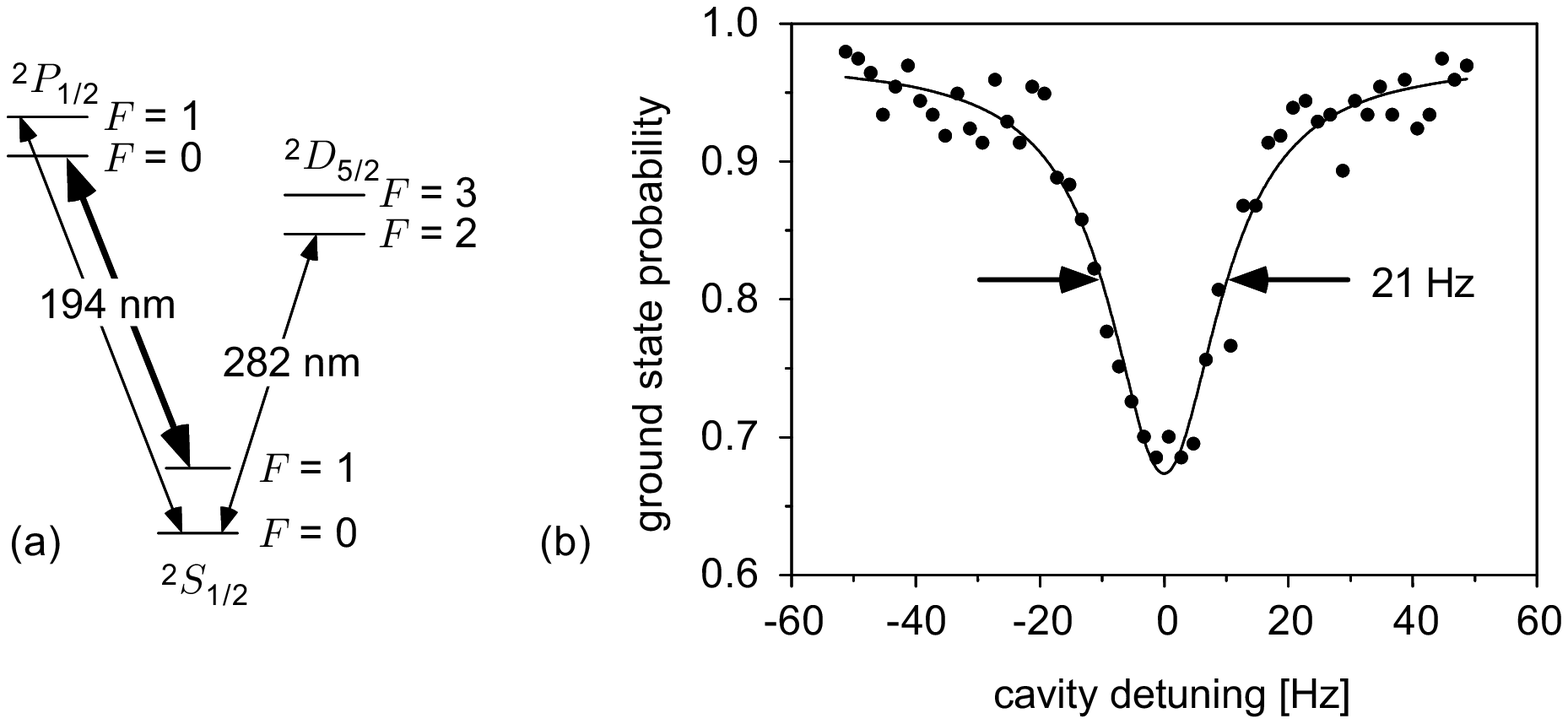}\vspace{1cm}
\vspace{0.3cm} \caption[]{(a) Partial level scheme for
${\rm^{199}Hg^+}$. The $194\;\mbox{nm}$ radiation is used for
Doppler cooling, state preparation and detection. The
$282\;\mbox{nm}$ transition from the ground state
$^2S_{1/2}(F=0,M_F=0)$ to the metastable $^2D_{5/2}(F=2,M_F=0)$
state provides the reference for the optical clock frequency. (b) A
typical spectrum of the $282\;\mbox{nm}$ clock transition obtained
under lock conditions is shown. Here, the excitation pulse length
was $20\;\mbox{ms}$, and the measured linewidth is Fourier transform
limited to about $20\;\mbox{Hz}$ at $563\;\mbox{nm}$
($40\;\mbox{Hz}$ at $282\;\mbox{nm}$).} \label{hg}
\end{center}
\end{figure}

The ${\rm Hg^+}$ and Ca systems have recently been described
elsewhere\cite{Rafac,Oates00,Oates99,Young}, so we summarize only
the basic features. The heart of the mercury optical frequency
standard is a single, laser-cooled ${\rm^{199}Hg^+}$ ion that is
stored in a cryogenic, radio frequency spherical Paul trap. The
$^2S_{1/2}(F=0,M_F=0)\leftrightarrow {^2D_{5/2}}(F=2,M_F=0)$
electric-quadrupole transition at $282\;\mbox{nm}$
[Fig.~\ref{hg}(a)] provides the reference for the optical standard
\cite{Rafac}. We lock the frequency-doubled output of a
well-stabilized $563\;\mbox{nm}$ dye laser to the center of the
quadrupole resonance by irradiating the $\rm{Hg^+}$ ion alternately
at two frequencies near the maximum slope of the resonance signal
and on opposite sides of its center. Transitions to the metastable
$^2D_{5/2}$ state are detected with near unit efficiency since the
absorption of a single $282\;\mbox{nm}$ photon suppresses the
scattering of many $194\;\mbox{nm}$ photons on the strongly allowed
$^2S_{1/2}-^2P_{1/2}$ transition\cite{Dehmelt,Bergquist87}. Usually,
48 measurements are made on each side of the resonance prior to
correcting the average frequency of the 282 nm source. If an
asymmetry between the number of excitations detected on the high-
and low-frequency sides is found, then the frequency of the probe
radiation is adjusted to minimize the asymmetry. In this way, we
steer the frequency of the $282\;\mbox{nm}$ source to the center of
the $S-D$ quadrupole resonance with an imprecision that decreases as
the square root of the measurement time. In Fig.~\ref{hg}(b) we show
an example of a normalized spectrum that was obtained from multiple
bidirectional scans through the resonance during the lock-up, where
the probe time was $20\;\mbox{ms}$. Most often, the frequency was
locked to resonance with a $10\;\mbox{ms}$ interrogation period,
which provided a fractional frequency instability reaching $~3 \times
10^{-15}\tau^{-1/2}$ for an averaging time $\tau$ measured in
seconds~\cite{Wineland}.

\begin{figure}[bthp]
\begin{center}
\includegraphics[width=8cm]{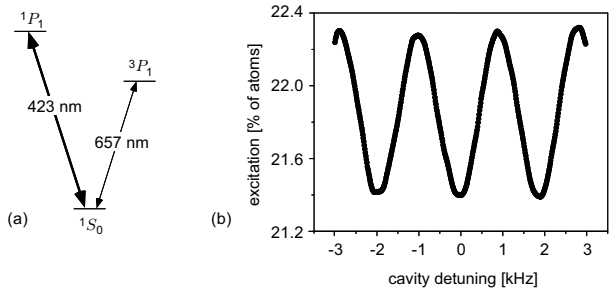}\vspace{1cm}
\caption[]{(a) Simplied diagram of the relevant energy levels in the
Ca standard. (b) Optical Bord\'{e}-Ramsey fringes with a
$960\;\mbox{Hz}$ (FWHM) resolution. The total averaging time to
generate this figure was $20\;\mbox{s}$.} \label{ca}
\end{center}
\end{figure}

The calcium standard is based on a collection of $\sim 10^7$
laser-cooled ${\rm^{40}Ca}$ atoms held in a magnetooptic trap. The
$423\;\mbox{nm}$ $^1S_0 \leftrightarrow {^1P_1}$ transition is used
for Doppler cooling and trapping the atoms to a residual temperature
of $\sim 2\;\mbox{mK}$, while the $657\;\mbox{nm}$ $^1S_0 (M_J=0)
\leftrightarrow {^3P_1}(M_J=0)$ clock transition ($400\;\mbox{Hz}$
natural linewidth) is used for the frequency standard
[Fig~\ref{ca}(a)]. We excite the clock transition with a four-pulse
Bord\'{e}-Ramsey sequence (pulse duration = $1.5\;\mbox{$\mu$s}$)
with light from a continuous wave (CW) frequency-stabilized diode
laser. Using a shelving detection technique similar to that employed
in the ${\rm Hg^+}$ system, near-resonant $423\;\mbox{nm}$ pulses
($5\;\mbox{$\mu$s}$ duration) are used before and after the
$657\;\mbox{nm}$ excitation to determine the fraction of atoms
transferred from the ground state. Figure~\ref{ca}(b) shows
Bord\'{e}-Ramsey fringes taken at a resolution of $960\;\mbox{Hz}$.
This system has demonstrated a fractional frequency instability of
$4 \times 10^{-15} \tau^{-1/2}$, when probing sub-kilohertz
linewidths~\cite{Oates00}. For the measurements presented here the
Ca spectrometer was operated with linewidths ranging from 0.96 to
$11.55\;\mbox{kHz}$ which are integer subharmonics of the recoil
splitting.

The recent introduction of mode-locked lasers to optical frequency
metrology greatly simplifies the task of optical frequency
measurements~\cite{Diddams,Jones,Holzwarth,Udem1,Reichert1,Niering}.
The spectrum emitted by a mode locked laser consists of a comb of
regular spaced continuous waves that are separated by the pulse
repetition rate $f_r$. The frequency of the ${n^{\rm th}}$ mode of
the comb is given by $f_n=nf_r+f_o$~\cite{Ferguson,Reichert2} where
$f_o$ is the frequency offset common to all modes that is caused by
the difference between the group- and the phase-velocity inside the
laser cavity. Whereas $f_r$ can be measured by direct detection of
the laser output with a photodiode, $f_o$ is measured by
heterodyning the harmonic of a mode $f_n=nf_r+f_o$ from the infrared
wing of the comb with a mode $f_{2n}=2nf_r+f_o$ from the blue side
of the comb~\cite{Jones,Holzwarth}. While an octave spanning comb
can be produced directly from a mode-locked laser~\cite{Kaertner},
launching the longer pulses from a commercially-available
femtosecond laser into an air-silica microstructure
fiber~\cite{pcf1,pcf2} also produces a frequency comb that spans an
octave.  Via nonlinear processes in the fiber, additional equally
spaced and phase-coherent modes are added to the comb. It has been
demonstrated that this process of spectral broadening preserves the
uniformity of spacing and spectral fidelity of the comb to at least
a few parts in $10^{16}$~\cite{Holzwarth}.

We couple approximately $200\;\mbox{mW}$ average power from a
femtosecond Ti:sapphire ring
laser($f_r\approx1\;\mbox{GHz}$)\cite{Bartels} into a
$15\;\mbox{cm}$ piece of a microstructure fiber that has a
$1.7\;\mbox{$\mu$m}$ core and a group velocity dispersion that
vanishes near $770\;\mbox{nm}$\cite{pcf1}. This power is sufficient
to increase the spectral width of the laser from $13\;\mbox{THz}$ to
more than $300\;\mbox{THz}$, spanning from $\sim 520\;\mbox{nm}$ to
$\sim 1170\;\mbox{nm}$. The infrared part of the comb from the fiber
($\lambda\approx1060\;\mbox{nm}$) is split off by a dichroic mirror
and frequency doubled into the green portion of the visible spectrum
with a $2\;\mbox{mm}$ long ${\rm KNbO_3}$ crystal. Following an
adjustable delay line that matches the optical path lengths, the
frequency-doubled light is spatially combined with the green part of
the original comb using a polarizing beam splitter. A second
rotatable polarizer projects the polarization of the combined beams
onto a common axis so that they can interfere on a photodiode. This
polarizer is also used to adjust the relative power of the two beams
for optimum signal-to-noise ratio (SNR) in the heterodyne signal. A
small grating prior to the photodiode helps to select only that part
of the frequency comb that matches the frequency-doubled light,
thereby reducing noise from unwanted comb lines~\cite{Reichert2}. We
phase-lock both $f_o$ and $f_r$ to synthesized frequencies derived
from a cavity-tuned hydrogen maser. Control of $f_r$ is achieved
with a cavity folding mirror that is mounted on a piezo transducer,
while $f_o$ is controlled by adjusting the $532\;\mbox{nm}$ pump
beam intensity with an electro-optic modulator~\cite{Holzwarth}.
When $f_o$ and $f_r$ are both phase-locked, the frequency of every
mode in the comb is known with the same precision as the reference
maser.

The CW light from the ${\rm Hg^+}$ ($563\;\mbox{nm}$) and Ca
($657\;\mbox{nm}$) spectrometers is transferred to the mode-locked
laser system via two single mode optical fibers that are
$130\;\mbox{m}$ and $10\;\mbox{m}$ long, respectively.
Approximately $2\;\mbox{mW}$ of CW light from each fiber is
mode-matched with the appropriate spectral region of the frequency
comb to generate a beat signal $f_b$ with a nearby mode. This beat
note is amplified and measured with a radio frequency counter. The
optical frequency is then expressed as $f_{opt}=f_o+mf_r+f_b$,
where $m$ is a large integer uniquely determined for each system
from previous coarse measurements of $f_{opt}$. 

We detect cycle slips in both of the phase-locks by monitoring
$f_r$ and $f_o$ with additional counters~\cite{Udem2}. We
selectively discard any measurement of $f_{opt}$ for which the
measured $f_o$ or $f_r$ deviate from the expected value by more
than $1/\tau_{gate}$, where $\tau_{gate}$ is the counter gate time
in seconds. We avoid miscounts of $f_b$ by using an auxiliary
counter to record the ratio $r$ between $f_b$ and $f_b/4$, where
the division by 4 is implemented digitally. Any measurements of
$f_b$ where the auxiliary counter gives a result that does not
satisfy $(r-4)*f_b<10/\tau_{gate}$ are discarded. We rely on the
assumption that the two counters recording $f_b$ and $r$, if in
disagreement, do not make the {\em same} mistake. For each data
point the three additional counters ($f_r$, $f_o$ and $r$) are
started before the counting of $f_b$, and operated with
$50\;\mbox{ms}$ longer gate times to ensure temporal overlap.

\begin{figure}[t]
\begin{center}
\includegraphics[width=8cm]{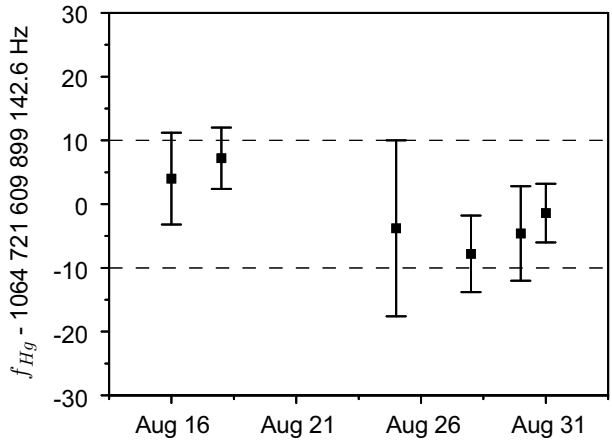}\vspace{1cm}
\vspace{0.5cm} \caption[]{A chronological record of the average
daily frequency of the ${\rm^{199}Hg^+}$ clock transition measured
on six days over a 15 day period representing $21\;651\;\mbox{s}$ of
total measurement time. The error bars represent statistical
fluctuations. The dashed lines represent an estimated systematic
uncertainty of $\pm10$ Hz in the ${\rm Hg^+}$ system in the absence
of a full evaluation.} \label{hgresult}
\end{center}
\end{figure}

\begin{figure}[t]
\begin{center}
\includegraphics[width=8cm]{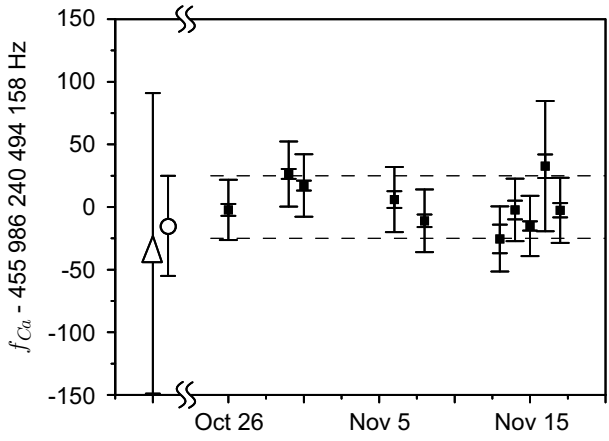}\vspace{1cm}
\vspace{0.5cm} \caption[]{The filled squares are the measured Ca
frequencies on ten days over a 23-day period representing $38\;787$
s of total measurement time. The inner and outer error bars for each
day represent the statistical and total uncertainties, respectively.
The dashed-lines show the $26\;\mbox{Hz}$ systematic uncertainty
assigned to the mean. The open triangle is the PTB measurement
reported in Ref.~\cite{PTB}, and the open circle is the Ca frequency
calculated from the present ${\rm Hg^+}$ result and our previous
measurement of the $76\;\mbox{THz}$ gap between Ca and ${\rm
Hg^+}$~\cite{Vogel01}.} \label{caresult}
\end{center}
\end{figure}

Figure~\ref{hgresult} summarizes the frequency measurements of ${\rm
Hg^+}$ made between Aug. 16 and Aug 31, 2000, and
Fig.~\ref{caresult} summarizes the Ca measurements made from Oct. 26
to Nov. 17, 2000. All measurements are corrected daily for the
second-order Zeeman shift and for the offset of the reference maser
frequency. The uncertainty for the Zeeman correction is
$<1\times10^{-15}$ for the ${\rm Hg^+}$ system and
$<2.5\times10^{-15}$ in the Ca system. The frequency of the maser is
calibrated by comparing to the local NIST time scale (5 hydrogen
masers and 3 commercial cesium clocks), which in turn is calibrated
by the local cesium fountain standard (NIST-F1~\cite{Jefferts}) as
well as international cesium standards. This resulted in a
fractional uncertainty in the frequency of the reference maser of
about $1.8\times 10^{-15}$ for the measurements.

The weighted mean of our measurements of the ${\rm Hg^+}$ clock
transition is $f_{Hg}=1\;064\;721\;609\;899\;143\;\mbox{Hz}$,
where the statistical uncertainty of $2.4\;\mbox{Hz}$ is near the
fractional frequency instability of the reference maser ($\sim 2
\times 10^{-13}$ at 1s, decreasing to $\sim 4 \times 10^{-16}$ at
a few days). Since we have not made a full evaluation of the ${\rm
Hg^+}$-standard, we only estimate the total systematic uncertainty
to be $10\;\mbox{Hz}$. The dominant systematic contribution to the
uncertainty of the $S-D$ transition frequency is believed to be
the electric-quadrupole shift of the $^2D_{5/2}$ state arising
from coupling with the static potentials of the trap. In our
spherical Paul trap, where the confinement of the ion uses no
static applied fields, the maximum quadrupole shift should be less
than $1\;\mbox{Hz}$ (or a fractional frequency shift
$<10^{-15}$)\cite{Itano}. In principle, it is possible to
eliminate the quadrupole shift by averaging the $S-D$ transition
frequencies for three mutually orthogonal orientations of a
quantizing magnetic field of constant magnitude. In the present
experiment, we have measured the $S-D$ frequency for various field
values, but we have made no attempt to eliminate the quadrupole
shift by using three orthogonal fields of constant magnitude. No
shift of the resonance frequency is observed within the precision
of these measurements. We anticipate that the uncertainties of all
systematic shifts in the ${\rm Hg^+}$ system can be reduced to
values approaching $1 \times 10^{-18}$~\cite{Rafac,Itano}.

For the Ca data shown [Fig.~\ref{caresult}], an additional
correction is applied each day to account for a frequency shift
caused by residual phase chirping on the optical Ramsey pulses
produced by amplitude modulating an acoustooptic modulator (AOM).
The phase chirping produced a resolution dependent frequency shift on
the order of $100\;\mbox{Hz}$ for $11.5\;\mbox{kHz}$ wide fringes but
only $10\;\mbox{Hz}$ for $0.96\;\mbox{kHz}$ wide fringes. On each
day, the Ca frequency was measured for $\sim 30$ minutes at each of
several fringe resolutions, and the zero-intercept of a linear fit
to the data was used as the corrected frequency. On the last 3 days
of measurements, we were able to reduce this shift by a factor of
$\sim3$ with improvements to the RF pulses that drive the AOM's. The
statistical uncertainty for each day's measurement (typically
$8\;\mbox{Hz}$) is smaller than the uncontrolled systematic
uncertainties in the the Ca frequency. The largest systematic
uncertainty stems from incomplete knowledge of the angular overlap
of the counterpropagating beams in the Ca spectrometer, combined
with a transverse drift velocity of the cold Ca ensemble. This leads
to a residual first-order Doppler shift with a magnitude $< 15$ Hz
(except on Nov. 16 where a large drift velocity led to a $\sim 52$
Hz uncertainty). Other significant uncertainties include our lack of
knowledge or control of electronic offsets and baseline asymmetries
($<12\;\mbox{Hz}$), wavefront curvature ($<10\;\mbox{Hz}$), and
cold-atom collisional shifts ($<10\;\mbox{Hz}$). Taking all known
systematic uncertainties in quadrature gives a confidence level of
$\sim 26\;\mbox{Hz}$ for the measured mean value as indicated by the
dashed lines in Fig.~\ref{caresult}.

Figure~\ref{caresult} also shows the good agreement between our
measurement and the most recent value measured with a harmonic
frequency chain~\cite{PTB}, which provides a degree of confidence in
the reproducibility of the Ca standards. 
An additional measure of the Ca frequency
can be made by using the present absolute measurement of ${\rm
Hg^+}$ and our earlier measurement of the
$76\;374\;564\;455\;429(40)\;\mbox{Hz}$ gap between $f_{Hg}/2$ the
and Ca standard~\cite{Vogel01}. This yields a value
$f_{Ca}=455\;986\;240\;494\;143(40)\;\mbox{Hz}$ in good agreement
with the present absolute measurement of $f_{Ca}$.

Finally, these results also provide data on the relative time
variability of atomic frequencies. S.~Karshenboim has recently
reviewed the implications of such comparisons and their contribution
toward constraining the possible time variation of fundamental
constants~\cite{Karshenboim}. In this regard ${\rm Hg^+}$ and Ca are
two of the most interesting cases to study. Comparing our present
measurement of $f_{Ca}$ to measurements made by PTB in
1997\cite{PTB} gives $(\partial f_{Ca}/\partial t)/f_{Ca}=(+2\pm
8)\times 10^{-14}\;{\rm yr^{-1}}$. Similarly, combining this result
with our May 2000 measurement of $f_{Hg}$ with respect to
$f_{Ca}$\cite{Vogel01} provides an initial baseline constraint on
the time variation of $f_{Hg}$ of $(\partial f_{Hg}/\partial
t)/f_{Hg}=(-7\pm 30)\times 10^{-14}\;{\rm yr^{-1}}$. Here we use the
defined unit of time based on the frequency of the Cs hyperfine
interval and assume that any time dependence is slow and dominantly
linear over the relevant time scale. At our present level of
precision we find no evidence of any relative time variation between
these three frequency standards, two optical and one microwave.

\vspace{0.5cm} The authors are most grateful to T.~Parker for
providing the crucial maser calibration and to A.~Bartels of
GigaOptics for his valuable assistance with the femtosecond laser.
We are also indebted to R.~Windeler of Lucent Technologies for
providing the microstructure optical fiber. We further acknowledge
many illuminating discussions with D.~Wineland, J.~Hall,
S.~Karshenboim, and F. Walls. This research was partially
supported by the Office of Naval Research and through a
Cooperative Research and Development Agreement with Timing
Solutions, Inc., Boulder, CO.

\vspace{0.5cm}

*Present address: Max-Planck-In\-sti\-tut f\"ur Quan\-ten\-op\-tik,
Hans-Kop\-fer\-mann-Str.1, 85748 Gar\-ching/Ger\-ma\-ny.

\end{document}